\pdfoutput=1

\documentclass[poms,final,nonblindrev]{poms1_V1} 

\usepackage{multirow}

\usepackage{bbm}
\usepackage{arydshln}

\usepackage{graphicx}
\usepackage{subfigure} 
\usepackage{natbib}

\usepackage{booktabs,caption}
\usepackage[flushleft]{threeparttable}

 \bibpunct[, ]{(}{)}{,}{a}{}{,}%
 \def\bibfont{\small}%

\usepackage[normalem]{ulem}
\usepackage{amsmath}
\newcommand{\stkout}[1]{\ifmmode\text{\sout{\ensuremath{#1}}}\else\sout{#1}\fi}

\TheoremsNumberedThrough     
\ECRepeatTheorems

\EquationsNumberedThrough    

\usepackage{booktabs} 
\usepackage[table]{xcolor}
\usepackage{enumitem}
\usepackage{graphicx}
\usepackage{amsmath}
\usepackage{booktabs}
\usepackage{tikz}
\usetikzlibrary{calc}

\newcommand\tikzmark[1]{%
  \tikz[overlay,remember picture] \coordinate (#1);}

\newcolumntype{L}[1]{>{\raggedright\let\newline\\\arraybackslash\hspace{0pt}}m{#1}}

\newcounter{marginparcounter}

\begin{document}

\thispagestyle{empty}  
\begin{center}
    \vspace*{0.5in} 
    
    \Large
    \textbf{Customer Service Operations: A Gatekeeper Framework}
    
	\vspace{0.5in}

\large	    
Maqbool Dada (Corresponding Author)\\
Carey Business School \\
Johns Hopkins University \\
100 International Drive \\
Baltimore, MD, 21202 \\
(765) 427-0751 \\
\EMAIL{dada@jhu.edu}

\vspace{0.5in}

Brett Hathaway \\
Marriott School of Business \\
Brigham Young University \\
730 TNRB \\
Provo, UT, 84602 \\
(801) 837-0474 \\
\EMAIL{brett\_hathaway@byu.edu}

\vspace{0.5in}

Evgeny Kagan \\
Carey Business School \\
Johns Hopkins University \\
100 International Drive \\
Baltimore, MD, 21202 \\
(734) 604-7152 \\
\EMAIL{ekagan@jhu.edu}

\end{center}

\newpage
\setcounter{page}{1} 
\pagestyle{plain} 


 \RUNAUTHOR{Dada, Hathaway, and Kagan}
\RUNAUTHOR{}

\RUNTITLE{Customer Service Operations}

\TITLE{Customer Service Operations: A Gatekeeper Framework}

\ARTICLEAUTHORS{%
\AUTHOR{Maqbool Dada}
\AFF{Carey Business School, Johns Hopkins University}
\AUTHOR{Brett Hathaway}
\AFF{Marriott Business School, Brigham Young University}
\AUTHOR{Evgeny Kagan}
\AFF{Carey Business School, Johns Hopkins University}
} 

\ABSTRACT{Customer service has evolved beyond in-person visits and phone calls to include live chat, AI chatbots and social media, among other contact options. Service providers typically refer to these contact modalities as ``channels''. Within each channel, customer service agents are tasked with managing and resolving a stream of inbound service requests. Each request involves milestones where the agent must decide whether to keep assisting the customer or to transfer them to a more skilled -- and often costlier -- provider. To understand how this request resolution process should be managed, we develop a model in which each channel is represented as a gatekeeper system and characterize the structure of the optimal request resolution policy. We then turn to the broader question of the firm's  customer service design, which includes the strategic problem of which channels to deploy, the tactical questions of at what level to staff the live-agent channel and to what extent to train an AI chatbot, and the operational question of how to control the live-agent channel. Examining the interplay between strategic, tactical, and operational decisions through numerical methods, we show, among other insights, that service quality can be improved, rather than diminished, by chatbot implementation.}%

\HISTORY{Received: May 2024; accepted: January 2025 by Michael Pinedo after one revision.}

\maketitle

%


\section{Introduction} \label{sec:Introduction}
The customer service channels available today vary along many dimensions, including synchronicity
(immediate interaction with a chatbot or live agent vs. delayed resolution via email or social
media), modality (text-based vs. voice), and the type of service provider (ranging from self-service
options like FAQs to live human agents to AI-driven chatbots). Digital channels, particularly
chatbots and social media, have seen a significant increase in adoption, with reports indicating a
double-digit adoption increase from 2018 to 2020 \citep{salesforce2020}. These channels often meet the
increased customer expectations for fast and flexible support, while helping reduce call volumes
and lower operational costs. At the same time, customers may resent dealing with an automated
technology unable to provide a more personal service experience \citep{aircall2023}. This tension presents firms with several strategic and operational challenges, as they seek to leverage the cost savings offered by service automation while maintaining high levels of customer satisfaction and loyalty.

\begin{figure}[bt]
\begin{center}
	\caption{Typical Customer Service Desk Architecture Implemented by BlackBeltHelp}\label{fig:blackbelt}
	\includegraphics[trim = 0in 0in 0in 0in, clip, width=6in]{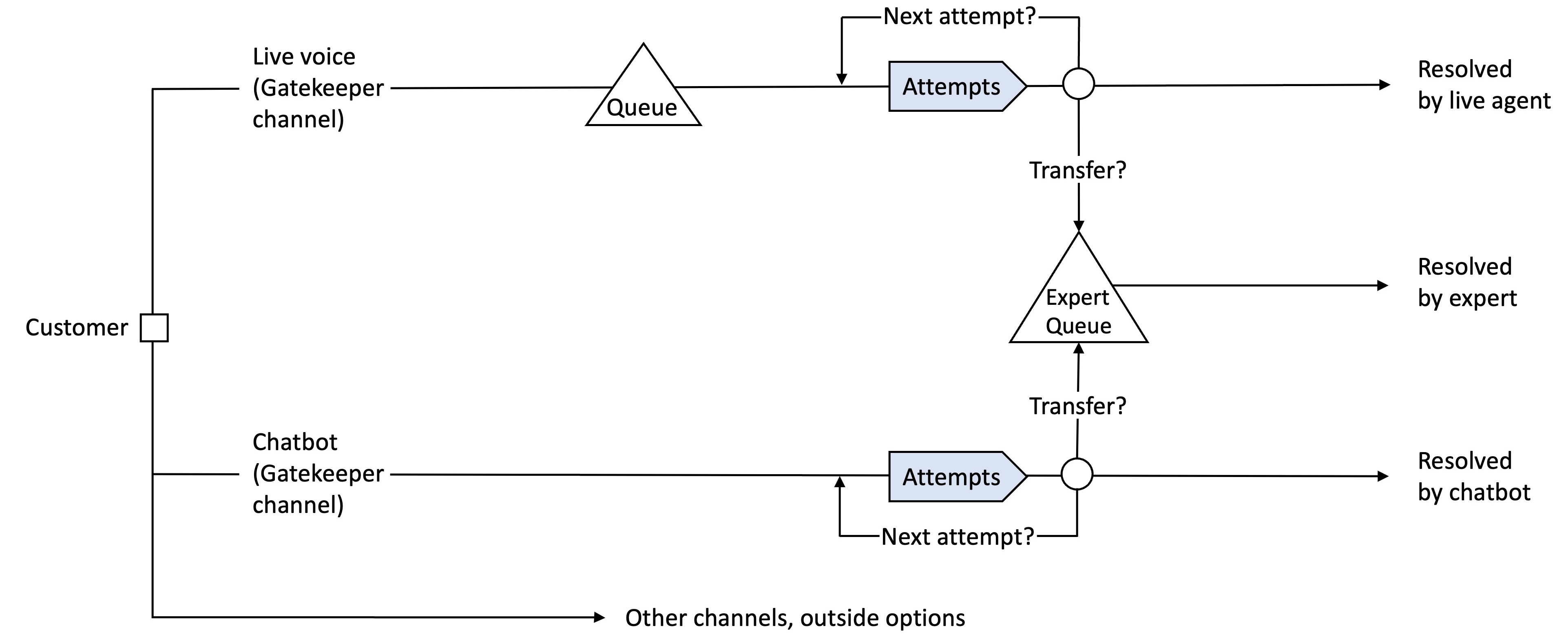}
\end{center}
\end{figure}

To better understand how multiple channels are managed in the customer service setting, we approached BlackBeltHelp
, a third-party provider of customer service solutions for the higher education sector. We interviewed BlackBeltHelp executives and engineers in November 2022. The following insights emerged from our conversations:

\begin{itemize}
\item \textbf{Multichannel architecture:} Different BlackBeltHelp clients choose different channel architectures. Some clients opt only for live channels. Other clients opt for a mix of live and AI-driven channels. BlackBeltHelp even reported at least one client who opted only for the chatbot channel. The client chooses which of these channels to offer and BlackBeltHelp provides the necessary training, staffing, and software development to integrate the channels in one platform. A representation of a two-channel architecture commonly implemented by BlackBeltHelp is shown in Figure \ref{fig:blackbelt}.

\item \textbf{Gatekeeper layers:} Customer request resolution is rarely a one-shot process -- correctly identifying the type of request and attempting to resolve it is an iterative process that involves milestones or steps codified by the client. While BlackBeltHelp is able to handle the majority of customer requests, a subset of them are ultimately routed to the client (dubbed ``expert'' in Figure \ref{fig:blackbelt}). Thus, BlackBeltHelp employees and chatbots almost invariably serve as gatekeepers to the expert workers at the client organization. 

\item \textbf{Chatbot channel has a unique performance and cost structure:} Clients are increasingly choosing to include an AI chatbot channel into their architecture. For a fixed development cost (plus minimal maintenance costs), the bot is capable of handling a virtually unlimited number of customer inquiries. Rather than paying by resolved request, clients only pay for software engineers to train the chatbot. Training the chatbot involves feeding it selected types of requests from a question database, beginning with the most frequently encountered inquiries and gradually incorporating more complex or less common scenarios. 

\end{itemize}

\vspace{0.1cm}
\noindent Our interactions with BlackBeltHelp suggest a complex range of configurations for designing customer service -- which channels to offer, and how to staff, manage and integrate them. Our study focuses on developing an analytical framework to address these questions. In particular, we examine the strategic-level question of the overall system architecture (\textit{What is the right mix of channels?}), the tactical-level question of what capacity to provide each channel (\textit{How many live agents to provide and how much to spend on chatbot training}), as well as the operational-level question of within-channel service design (\textit{When and how should transfers be performed?}).

While the first two questions follow directly from our interactions with BlackBeltHelp and other customer service providers, the issues related to within-channel service design (transfer policies) are less well-defined. To better understand these issues, and the related trade-offs, we conducted an online  survey of customer attitudes towards transfers in customer service. The key takeaways of the survey are that both presence and the type of transfer matters. Specifically, the survey data suggest that customers prefer more personalized transfers (often referred to as ``warm transfers'' in industry), where the transferring agent provides context and passes on the relevant information to the receiving (expert) agent before leaving the call. Conversely, customers have a strong aversion to the more common ``cold transfer'', where the customer is transferred without any context or personalization. Indeed, the preference for ``warm transfers'' holds even when such transfers result in longer waiting times. See the electronic companion for survey details.

The insights from the survey motivate our operational control problem of optimal gatekeeping, i.e., the question of optimal transfer decisions within a single channel of the service architecture. In \textsection 3 we characterize service resolution as a sequential $S$-attempt process. Each of the $S$ resolution attempts has its own processing time and success probability. At the end of an attempt, if the customer issue has not been resolved, the agent handling the request chooses one of three actions: 1) \textit{continue} to the next attempt, 2) \textit{warm transfer}, wherein the agent accompanies the customer and facilitates the transition to the next channel, or 3) \textit{cold transfer}, wherein the agent drops the customer into the next channel.

The firm's objective is to determine optimal resolution policies for the above problem, while accounting for congestion. Specifically, we consider customer traffic (upstream congestion) and expert availability (downstream  congestion). Further, channels may have limited operating hours. To account for these factors, we formulate the decision problem as a finite-horizon stochastic dynamic program. The solution to this problem includes a carefully specified set of terminal conditions formulated in \textsection 4 that make the optimal resolution policy stationary, thereby accommodating both 24/7 operations and finite work shifts. Stationarity is then exploited to further structure the optimal policy, which is shown to depend on congestion and resolution attempt.

In \textsection 5, we explore heuristic solutions and present model extensions. We begin with numerical experiments that restrict the policy search to threshold policies, which have the advantage of being  computationally efficient and can be found in polynomial time. We also identify a simple sufficient condition—related to the shortest-processing-time (SPT) rule—for threshold policy optimality. We then incorporate queueing dynamics into the model by introducing a finite waiting room. Although the policy space expands considerably with this addition, we find that a heuristic policy of transferring customers when the waiting room reaches half capacity achieves near-optimal performance across a range of arrival intensities.

In \textsection 6 we turn to the broader problem of examining how the optimal operational control decisions interplay with the tactical and strategic customer service design questions faced by the firm. In particular, the firm's customer service design problem consists of the following components: 
\begin{itemize}
\item[]\textbf{Strategic}: Which gatekeeper channels (human, AI chatbot, both) to offer? 
\item[]\textbf{Tactical}: How should these channels be staffed and/or trained? How much should be spent on AI development?
\item[]\textbf{Operational}: What are the optimal resolution policies within each channel? 
\end{itemize}

To solve this joint problem, we first re-examine the gatekeeping problem when a chatbot, rather than a live agent, acts as a gatekeeper. Chatbot systems have a unique cost structure (fixed costs of development, rather than variable costs of staffing), face virtually no upstream congestion, and their actions are more limited compared to human gatekeepers, all of which significantly simplifies the problem. We then characterize the effects of the channel architecture, staffing, chatbot training and resolution policies on service quality, which in turn determine customer demand (arrivals to each channel) as a response to the channel-specific performance measures announced by the firm.

We conclude by presenting numerical examples of the optimal customer service design over various chatbot training cost and agent wage scenarios. Unsurprisingly, different combinations of chatbot training costs and live-agent wages can lead to different channel architectures being optimal: live-agent-only, chatbot-only, or both. There are also important interactions between optimal chatbot training, staffing levels and resolution policy. In particular, the introduction of a chatbot channel may not necessarily result in reduced staffing but may change the optimal live-agent resolution policy to one that delivers higher quality service. More broadly, our model demonstrates that when choosing how to implement AI chatbot technology, managers should consider not only the implications for  staffing but also the more indirect consequences for service policies within each channel.

Our work makes three key contributions. First, we develop a comprehensive model of optimal gatekeeping. While gatekeepers are becoming a more prevalent lens for studying service and healthcare systems \citep{shumsky2003,freeman2017,hathaway2022}, optimal gatekeeper behavior remains understudied. Our results bridge an important gap in theoretical development in this area. Second, we comprehensively examine the firm's customer service design problem, which includes channel selection, automated channel investment, channel staffing, and within-channel resolution policies. While components of this problem have been studied in isolation \citep{shumsky2003,hasija2005,gao2018,feldman2023,kagan2023}, this is the first attempt (that we are aware of) to integrate them within a single framework. Third, we examine the role of chatbots within a broader service operations framework. As chatbot adoption continues to rise \citep{cohen2018,sheehan2020,deloitte2023}, it is becoming increasingly important to develop analytical approaches to better understand their value (and limitations) in customer service.

\section{Overview and Related Literature} \label{sec:Empirical}
\subsection{Overview}
At a high level, the firm's customer service strategy includes choosing which service channels to offer, at what levels to staff any live channels (such as phone and live chat), to what extent to develop any automated/self-service channels (such as FAQs and AI chatbots), and what resolution policies to implement (such as how long to attempt to resolve a given customer issue and when and how to transfer customers within and across channels). If we restrict our attention to two focal channels, a live-agent channel and a chatbot channel, and denote their profit contributions by $\pi^{agent}$ and $\pi^{bot}$, then the firm's profit-maximization problem is given by: 
\begin{align} \label{eq:architecture}
\max\limits_{k^{agent},\boldsymbol{D^{agent}},p^{bot}_{succ}} \pi^{agent}(k^{agent},\boldsymbol{D^{agent}},\lambda^{agent}(k^{agent},\boldsymbol{D^{agent}},p^{bot}_{succ})) + \pi^{bot}(p^{bot}_{succ},\lambda^{bot}(k^{agent},\boldsymbol{D^{agent}},p^{bot}_{succ})), 
\end{align}

\noindent where $k^{agent}$ is the live-agent staffing level, $\lambda^{agent}$ and $\lambda^{bot}$ are the arrival rates to each channel, $\boldsymbol{D^{agent}}$ is the live-agent resolution policy, and $p^{bot}_{succ}$ is the chatbot success probability the firm has trained the bot up to. Note that the firm may choose to staff no live agents ($k^{agent}=0,\boldsymbol{D^{agent}}=0$) or not to employ a chatbot ($p^{bot}_{succ}=0$). Also note that because the bot can process a virtually unlimited number of requests, no staffing decision is required for the bot channel. Hence, the firm selects the optimal channels, the staffing and capability levels ($k^{agent}$,$p^{bot}_{succ}$), and how these channels should operate ($\boldsymbol{D^{agent}}$). Utility-maximizing customers observe these choices and respond by choosing which channel (if any) to join, with their collective decisions determining $\lambda^{agent}$ and $\lambda^{bot}$.

In the remainder of \textsection 2 we review research related to this problem. In \textsection 3 we introduce the operational problem of optimally setting $\boldsymbol{D^{agent}}$, i.e., the resolution policy for a channel staffed by a single live agent. In \textsection 4 we characterize the structure of the optimal policy for this problem. In \textsection 5 we examine heuristic solutions and extend the operational model to allow for queueing dynamics and multiple live agents. In \textsection 6 we show how the problem differs for the bot, i.e., how the firm should set $p^{bot}_{succ}$, examine the problem from the customer perspective by characterizing $\lambda^{agent}$ and $\lambda^{bot}$, and conclude with a numerical example that illustrates that the firm should jointly choose $k^{agent},\boldsymbol{D^{agent}}$, and $p^{bot}_{succ}$ while accounting for customer response to those choices.

\subsection{Related Research}
\subsubsection*{Channel Architecture}
Optimal channel mix problems have been studied in the retail context, where the integration of physical and digital channels has become common practice \citep{brynjolfsson2013,gao2017,gallino2019}. Multichannel strategies are especially common in the food service industry, where congestion plays a central role \citep{gao2018,feldman2023}. Multichannel environments have also been studied in marketing with a focus on understanding customer experiences in different channels \citep{ansari2008customer,lund2014managing,lemon2016understanding}. No studies that we are aware of focus specifically on channel selection in the customer service domain.

\subsubsection*{Resolution Process} The resolution policies and training levels ($\boldsymbol{D^{agent}},p^{bot}_{succ}$) set the protocols for managing incoming customer requests; in particular, how much time and effort the agent (or chatbot) should spend on each request, and how and when requests should be transferred to an expert. Each channel therefore acts as a gatekeeper system \citep{shumsky2003,hasija2005,lee2012}, in which simple requests are handled by lower-cost workers (or chatbots), while more complex tasks are reserved for higher-cost specialists or experts. \cite{hathaway2022} examine experimentally whether gatekeepers correctly incorporate incentives to cold transfer while accounting for upstream congestion present in such systems. \cite{freeman2017} and \cite{batt2017} study work-sharing behaviors in healthcare. Also related are \cite{alizamir2013} and \cite{kremer2023}, who examine optimal stopping in sequential diagnostic processes under congestion, and \cite{sun2022triage}, who study optimal triage under congestion. Our gatekeeping model differs from the existing research in that we (a) expand the action set to include multiple types of transfers, and (b) incorporate upstream and downstream congestion.

\subsubsection*{Channel Joining Behavior}
In \textsection 6 we model channel demand, $\lambda^{agent}$ and $\lambda^{bot}$, as an endogenous response to the firm's service offering. Our work is therefore related to the literature on optimal routing models that incorporate customer decisions \citep{armony2004,vericourt2005,ganszhou2007}, as well as to the  behavioral literature on queue joining \citep{naor1969,kremer2016,allon2018}. Gatekeeper systems, i.e., systems in which customers are frequently transferred, may make the wait more uncertain as well as more fragmented, both of which may reduce channel uptake \citep{leclerc1995,kumar2021}. Abrupt transitions within service experiences, which are more likely in gatekeeper systems, may be particularly memorable and can  negatively affect customer satisfaction and channel uptake \citep{lee2008, dasgupta2016}. Our model of customer channel joining behavior explicitly accounts for such factors.

\subsubsection*{AI Chatbots in Customer Service}
While AI-driven chatbots are becoming increasingly capable of solving complex problems, there remain significant barriers to their adoption in customer service. A key barrier is that chatbot interactions often result in a transfer, necessitating a complete restart of the service process with a human provider \citep{kagan2023}. Our model will refer to such transfers as \textit{cold} transfers, as opposed to more seamless and personalized \textit{warm} transfers, which chatbots are typically unable to provide \citep{deloitte2023}, and which customers prefer over cold transfers. Separately, customers may dislike interactions with an algorithmic (rather than human) service provider, and that may reduce their willingness to use the chatbot option \citep{dietvorst2015,luo2019,sheehan2020,kagan2023}.  Attitudes regarding algorithms are incorporated into our model of customer preferences via a technology aversion parameter, affecting channel demands $\lambda^{agent}$ and $\lambda^{bot}$ in (\ref{eq:architecture}).

\subsubsection*{Related Dynamic Programming Analyses}

Our model of the gatekeeper resolution process is most closely related to  \cite{shumsky2003}, who also model a single gatekeeper responding to a stream of incoming customers (patients). They focus on the question of optimal compensation structure, which, for practical reasons, is typically time- and congestion-state-invariant within a shift. Therefore, they use the average reward criterion \citep{puterman2014}, which assumes stationarity and significantly simplifies analysis. Because we are less interested in compensation and because modern technology enables agents to have high visibility into traffic and system congestion, we explicitly model congestion (both upstream and downstream) and time remaining in the shift. This allows us to develop more textured insights into optimal gatekeeper behavior.

Since agents work in finite shifts (which can be quite short due to the part-time nature of the work), our model requires careful specification of the terminal conditions in the dynamic program. The appropriate choice of terminal conditions is thus a critical technical aspect of our model. A theoretically sound remedy for finite-horizon representation of such Markov decision processes is provided by \cite{veinott1966}, who developed the ``buy-back'' assumption in inventory models. Under this assumption, the finite-horizon problem exhibits the same localized behavior as its infinite-horizon counterpart. Applications of Veinott’s approach can be found in the state-dependent inventory model of \cite{xu2017finite} and the fishing stocks management model of \cite{lovejoy1988effect}. We follow the Veinott approach and carefully specify a set of state-dependent linear terminal conditions, which accommodates both finite-horizon settings, such as shift operations, and round-the-clock settings, such as service call centers.

\section{Live-Agent Model} \label{sec:Live Agent Model} 
In this section we focus on the live-agent channel and develop a dynamic programming model for how the firm should set the resolution policy for this channel ($\boldsymbol{D^{agent}}$ in (1)). This channel is staffed by a number of agents who are trained to follow the same protocol determined by the firm. The admission control is organized such that each agent receives an equitable load, making the system scalable in the number of agents. Hence, it is convenient to simply model the channel as a single representative live agent receiving a stream of incoming customer requests. (In \textsection 6 we will discuss a multiserver extension of the model.) The agent receives a flat payment and acts in the interest of the firm. The agent can be an internal employee, a worker in an outsourced call center, or can be employed by a third-party provider such as BlackBeltHelp (\textsection 1). 

The key service process decision facing the firm  is to determine, given current congestion patterns, how long the agent should continue working with each customer, and when and how that customer should be transferred to an expert service provider. The trade-off from the firm's perspective is between allowing its agents to spend more time with each customer, leading to fewer transfers and higher quality service, vs. instructing its gatekeepers to transfer earlier (and do so in a cost-effective manner).

\subsection{Resolution Process} \label{subsec:resolution process}
The agent operates in a shift of $T$ discrete periods and is paid at a rate of $c^{wage}$ per period. At the beginning of each period $t=1,\cdots,T$, a customer arrives with probability $q$. If the agent is busy, since there is no waiting room, the customer is not admitted and leaves the system. If the agent is idle, has just successfully resolved a request, or just transferred a request, then the customer is admitted for service and service starts at $t+1$. If admitted, the customer remains in the system until their issue is resolved successfully. The firm receives revenue of $r$ for each admitted customer when the customer issue is resolved and the transaction is closed. Note that our dynamics are consistent with a queuing system with blocking and no reneging; in \textsection 5.2 we extend the model to include a finite waiting room.

For each admitted customer, as in \cite{alizamir2013} and \cite{kremer2023}, the agent sequentially executes solutions from a pre-ordered list determined by the firm, until either the issue is resolved or the customer is transferred.  Specifically, the resolution process is represented by an ordered list of potential solutions $s \in \{1,2,..., S\}$, where each attempt $s$ is characterized by its handling time, $\tau_s$, and the conditional probability that it resolves the request, $\rho_s$, which depends on the number of previously unsuccessful attempts (since the order cannot be changed by the agent). Without loss of generality, one of the $S$ attempts is guaranteed to resolve the request so that $\rho_S =1$. The focal decision of the firm is \textit{how many} attempts the agent will make before transferring the customer to the expert. In particular, upon completion of attempt $s$, one of two outcomes may occur: (1) attempt $s$ resolves the request or (2) attempt $s$ does not resolve the request and the agent follows the firm's resolution policy by performing one of three actions:
\smallskip

\noindent \textbf{Continue:} The agent makes the next attempt ($s+1$). This expends $\tau_{s+1}$ time units, and resolves the request with probability $\rho_{s+1}$.
\smallskip

\noindent \textbf{Warm Transfer:} The agent warm transfers the customer to an expert. In practice, this may include introducing the customer to the expert, providing context, or passing on any relevant customer information to avoid repetition. A warm transfer costs the firm $c^w$ for expert compensation, where  $c^w$ represents a transfer payment to a different division of the same firm where the expert is employed, or a fee to the third party that employs the expert. The duration of the warm transfer depends on whether an expert is immediately available to receive the customer. In each period, an expert is available ($A = 1$) with probability $a$ and unavailable ($A=0$) with probability $1-a$. Once the warm transfer begins, the agent expends $\tau_w$ periods performing the transfer. Following the warm transfer, the agent remains idle if there is no arrival ($Q = 0$), or begins handling a new request if there is an arrival ($Q=1$).

\vspace{.05in}

\noindent \textbf{Cold Transfer:} The agent cold transfers the customer to an expert, who resolves the request. A cold transfer costs the firm $c^c$ for expert compensation, where $c^c > c^w$ since the agent simply drops the customer into the expert queue, requiring the expert to spend more time with the customer. If there is no arrival ($Q = 0$), the agent is idle until another request arrives; otherwise, there is an arrival ($Q = 1$) and the agent immediately begins handling a new request.  

\smallskip

For each request the decision repeats until it is either resolved or transferred to the expert. After each resolved (or transferred) request, the agent begins serving the next customer as soon as one becomes available. Both congestion states ($Q$ and $A$) are observable. Thus, the key trade-off at each decision point is whether, given current congestion, to dedicate more agent time to the focal customer to increase the chances of resolving their request, or to process a higher volume of customers by referring them to the expert, thereby freeing up agent capacity.

\subsection{Dynamic Programming (DP) Formulation} \label{subsubsec:DP Formulation}
To formulate the DP for the above problem, we need to define the state of the system in period $t = 1,\cdots,T$.  To that end, let $X \in \{0^i,0^w,1,2,\cdots,S\}$ denote the agent's status, where $X\in\{1,2,\cdots, S\}$ indicates that the agent just finished attempt $s$,  $X=0^i$ indicates an auxiliary state in which the agent was idle in the previous period waiting for a new request to process, and   $X=0^w$ indicates the other auxiliary state in which the agent has initiated a warm transfer but was waiting with the customer in the previous period because no expert was available to receive the customer. Thus, the state of the system is fully specified by the 3-tuple $(X,Q,A)$. At time $t$, the objective is to choose the action that maximizes expected net revenue over time $t$ to $T$, exclusive of the ``sunk'' agent wage, which need not be considered. Denote the optimal value function at $t = 1,\cdots,T$ by $V_t(X,Q,A)$, and the terminal value vector by $V_{T+1}(X,Q,A)$. 

Before writing the value function for each of the $4(S + 2)$ states, we make two observations. First, if the agent was idle waiting for a new request to arrive in the previous period ($X=0^i$), then the availability of the expert is irrelevant since the agent will either continue to be idle (if $Q=0$) or begin working on the first attempt of a new request (if $Q=1$). Hence, the value function does not depend on $A$ and can be conveniently denoted by $V_t(0^i,Q,-)$. Second, if the agent has initiated a warm transfer and was waiting for an expert to become available ($X=0^w$), then the presence of an arrival is irrelevant since the agent will either continue to wait for the expert (if $A=0$) or begin handing the request off to the expert (if $A=1$). Hence, the value function does not depend on $Q$ and can be conveniently denoted by $V_t(0^w,-,A)$. 

These simplifications allow us to write the value function for $t = 1,\cdots,T$ as follows: 
$$ V_t(X,Q,A)= \begin{cases}
(1-q)V_{t+1}(0^i,0,-) + qV_{t+1}(0^i,1,-),\hspace{1.53in}\text{if $X=0^i, Q=0, \hspace{.08in}(2.1)$}\\

(1-q)(1-a)V_{t+\tau_1}(1,0,0) + q(1-a)V_{t+\tau_1}(1,1,0) + \cdots \\

\hspace{.1in} (1-q)aV_{t+\tau_1}(1,0,1) + qaV_{t+\tau_1}(1,1,1), \hspace{1.34in} \text{if $X=0^i, Q=1, \hspace{.08in}(2.2)$}\\

(1-a)V_{t+1}(0^w,-,0) + aV_{t+1}(0^w,-,1),\hspace{1.45in}\text{if $X=0^w, A=0, \hspace{.06in}(2.3)$}\\

(1-q)V_{t+\tau_w}(0^i,0,-) + qV_{t+\tau_w}(0^i,1,-),\hspace{1.41in}\text{if $X=0^w, A=1, \hspace{.06in}(2.4)$}\\

\rho_X(r+V_t(0^i,Q,-))+ (1-\rho_X)\big[\max\{\mathcal{N}_t(X),\cdots \\

\hspace{.1in}-(c^w-r)+V_t(0^w,-,A), -(c^c-r)+V_t(0^i,Q,-)\}\big], \hspace{.53in} \text{if $1\leq X <S, \hspace{.30in}(2.5)$}\\

r+V_t(0^i,Q,A), \hspace{2.88in} \text{if $X=S, \hspace{.55in}(2.6)$} \\
\end{cases}$$

\stepcounter{equation}

\noindent where $\mathcal{N}_t(X)$ in (2.5) is the value-to-go for making attempt $X+1$ (continuing) after attempt $X$ fails in period $t$. This is given by $$ \mathcal{N}_t(X) = (1-q)(1-a)V_{t+{\tau_{X+1}}}(X+1,0,0) + q(1-a)V_{t+{\tau_{X+1}}}(X+1,1,0) + \cdots $$ \begin{equation} \label{eq:VC} (1-q)aV_{t+{\tau_{X+1}}}(X+1,0,1) + qaV_{t+{\tau_{X+1}}}(X+1,1,1).\end{equation} In (\ref{eq:VC}) each of the four terms on the right side of the inequality is the probability of being in a given congestion state $(Q,A)$ after completing attempt $X+1$, multiplied by the value-to-go of being in that congestion state upon completion.

Equation $(2.1)$ assures that if the agent was idle during period $t$ because there was no arrival, the agent begins period $t+1$ idle, whereas $(2.2)$ assures that if there was an arrival, the agent began service. Equation $(2.3)$ assures that if the agent was idle during period $t$ waiting for an expert, the agent begins period $t+1$ idle, whereas $(2.4)$ assures that if the expert became available, the agent spends $\tau_w$ periods performing the handoff and is idle at the beginning of period $t+\tau_w$. Equation $(2.6)$ guarantees that if the final potential solution was attempted ($X=S$), the firm received revenue $r$, and the agent moved to the next customer as soon as one became available. 

The key trade-off is shown in $(2.5).$ In this case, with (conditional) probability $\rho_X$, attempt $X=s$ resolves the request, the firm receives the revenue $r$ and the firm instructs the agent on how to proceed optimally. With complementary probability $1-\rho_X$, attempt $X=s$ did not resolve the request. So, to maximize expected future value, the agent must either make the next attempt $X+1$ (the first argument of the max operator), warm transfer and incur the warm transfer cost of $c^w$ (the second argument), or transfer the request and incur the cold transfer cost of $c^c$ (the third argument). Note that in all cases, the firm eventually receives revenue $r$ from the customer. Finally, we note that in our formulation, the agent may take an action that results in $t$ exceeding $T$. We handle this in our specification of $V_{T+1}(X,Q,A)$ (the terminal conditions), whose discussion we defer as they play a critical role in characterizing the structure of the optimal policy.

\begin{figure}[t]
\begin{center}
	\caption{Value-To-Go After Failed Attempt $X$ For Each Action And Congestion State $(Q,A)$}\label{fig:81}
	\includegraphics[trim = 0in 0in 0in 0in, clip, width=6in]{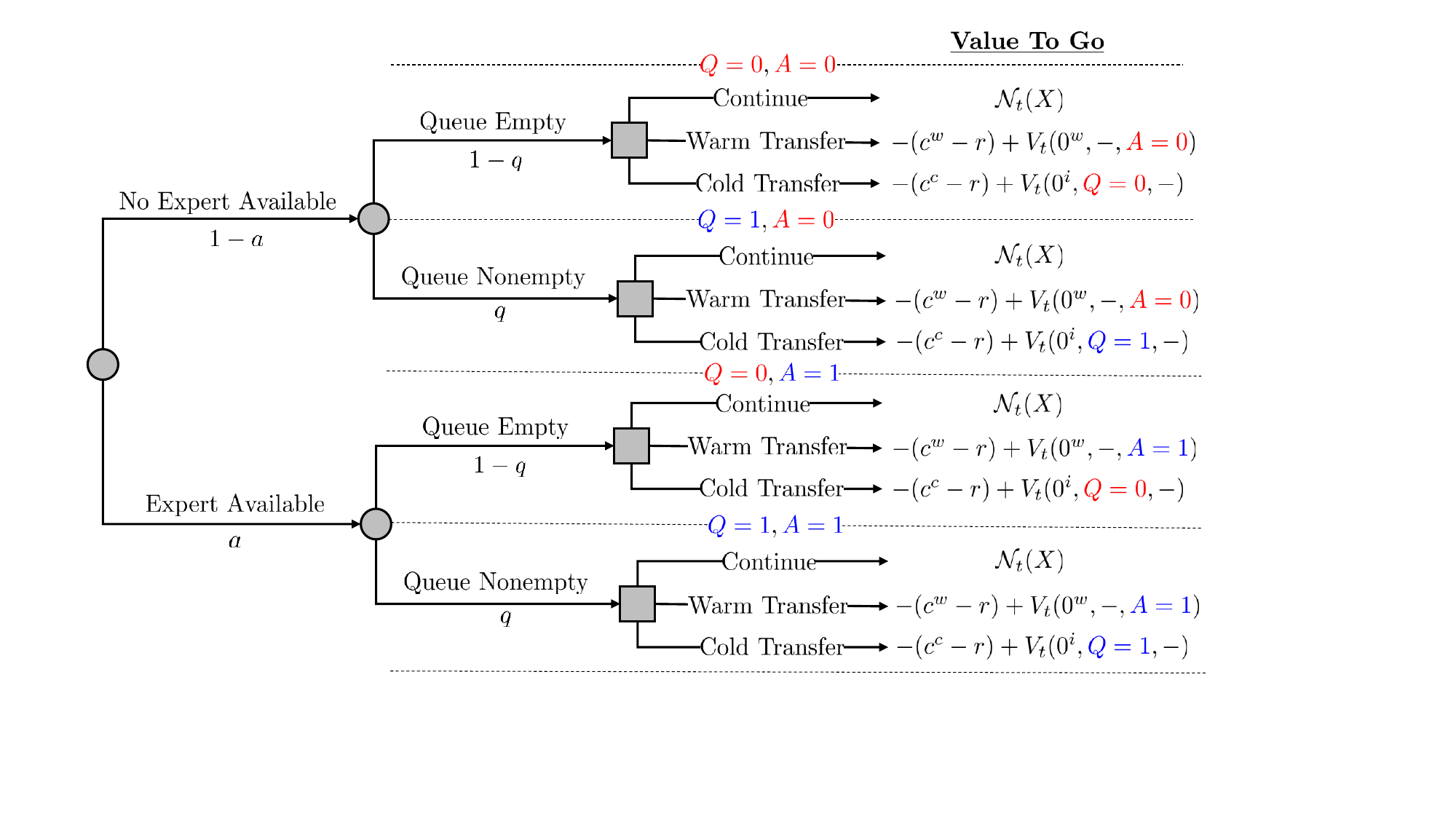}
\end{center}
\end{figure}

The essence of the problem is captured intuitively by Figure \ref{fig:81}, in which circles represent random nodes and squares represent decision nodes. The agent has just failed to resolve the request after making attempt $X$. The state is defined by downstream congestion ($A$) and the presence of an arriving customer ($Q$). For each $(Q,A)$ combination, one of the three actions is optimal: 1) continuing, resulting in the value-to-go of $\mathcal{N}_t(X)$, 2) warm transferring, resulting in a net revenue of $r - c^w$, and 3) cold transferring, resulting in a net revenue of $r - c^c$, with either transfer option leading to a transition to a new state.

Several observations are in order. First, because $Q$ and $A$ are time-independent, $\mathcal{N}_t(X)$ depends only on $X$, i.e., how many unsuccessful attempts have been completed. Thus, the value of continuing is constant across the congestion states. Second, to visualize upstream congestion, we highlight instances where $Q=1$ in blue and instances where $Q=0$ in red. When comparing the top two decisions (($Q=0,A=0$) and ($Q=1,A=0$)), the only difference between the value-to-go for each action is that it is greater for cold transferring when $Q=1$ than when $Q=0$, because cold transferring when there is no arrival leads to nonproductive idle time. The same holds when comparing the bottom two decisions. Third, to visualize downstream congestion, we highlight instances where $A=1$ in blue and instances where $A=0$ in red. Analogous to upstream congestion, when comparing congestion in the first and third decisions, the only difference between the value-to-go for each action is that it is greater for warm transferring when $A=1$ than when $A=0$, because when there is no expert available, warm transferring requires the agent to wait idly until one becomes available. The same holds when comparing the second and fourth decisions. We use these observations to derive local properties that help simplify the structure of the optimal policy.

\section{Policy Analysis} \label{sec:analysis} In this section we use the above dynamic programming formulation to provide structure to the optimal resolution policy. We first derive local properties that substantially reduce the number of admissible decision rules. We then propose a set of terminal conditions that make the optimal policy stationary over time. Finally, we identify optimal policy structure under stationarity.

\subsection{Local Properties} \label{subsec:Local Properties}
The dynamic program in (2.1) - (2.6) generates a large number of candidate policies. Since there are four possible states, each with a choice of three actions, at each $X$ and $t$, the strategy space consists of $3^4 = 81$ possible action vectors. Ordinarily, such a large number of possible policies would be daunting, even if the problem were stationary. Fortunately, we are able to reduce complexity by finding a set of dominance rules that eliminate all but 11 of the 81 decision rule vectors. The proof of this and the following two theorems are in the electronic companion.

\vspace{.1in}

\noindent \textbf{THEOREM 1A} \textit{If in period $t$, under a given $A$ it is optimal to warm transfer when there is an arrival ($Q=1$), then it is also optimal to warm transfer when there is no arrival ($Q=0$). If in period $t$, under a given $A$ it is optimal for the agent to continue when there is an arrival ($Q=1$), then it is also optimal to continue when there is no arrival ($Q=0$).}

\vspace{.1in}

\noindent \textbf{THEOREM 1B} \textit{If in period $t$, under a given $Q$ it is optimal to cold transfer when an expert is available ($A=1$), then it is also optimal to cold transfer when no expert is available ($A=0$). If in period $t$, under a given $Q$ it is optimal for the agent to continue when an expert is available ($A=1$), then it is also optimal to continue when no expert is available ($A=0$).}

\vspace{.1in}

\noindent Theorem 1A relies on the fact that neither the payoff of warm transferring nor the payoff of making the next attempt depends on the presence of an arrival ($Q$), and the payoff of cold transferring is lower when there is no arrival ($Q=0$). Analogously, Theorem 1B relies on the fact that neither the payoff of cold transferring nor the payoff of making the next attempt depends on current expert availability ($A$), and the payoff of warm transferring is lower when no expert is available ($A=0$).

Table \ref{tab:decision_rules} lists the admissible decision rules that remain after eliminating those that do not satisfy the dominance conditions in Theorems 1A and 1B. The first rule is C(ontinue), which is to continue with the next attempt irrespective of congestion state ($Q,A$). The next four are H(ybrid) rules, in which (depending on the state) the agent may continue, warm transfer, or cold transfer. The final six are T(ransfer) rules, in which (depending on the state) the agent always transfers (where the mode depends on the state). From the customer perspective (which we will discuss in \textsection 6) the rules also map to service quality. For example, rule C is the pure single-channel strategy (concierge service) as the customer remains with the agent for the next attempt, regardless of congestion. In contrast, rules T1 through T5 correspond to lower service quality and transfer the customer, with the transfer mode depending on the state. Among the T rules, T5 is the rule with the lowest service quality, as the customer is always cold transferred, regardless of state.

\begin{table}[h!]
  \caption{Admissible Decision Rules \label{tab:decision_rules}}
 \makebox[0.95\textwidth][r]{ \footnotesize
     \begin{tabular}{L{2.2cm}cccc}
    \toprule
    \multicolumn{1}{r}{\textbf{State:}} & $(Q=0, A=0)$ & $(Q=1, A=0)$ & $(Q=0, A=1)$ & $(Q=1, A=1)$ \\
       \textbf{Label} &   &   &   &  \\
    \midrule
    C  & Continue & Continue & Continue & Continue \tikzmark{start}  \\
    \hdashline
    H1    & Continue & Continue & Warm Transfer & Warm Transfer\\
    H2    & Continue & Cold Transfer & Warm Transfer & Warm Transfer \\
    H3    & Continue & Cold Transfer & Warm Transfer & Transfer \\    
    H4    & Continue & Cold Transfer & Continue & Cold Transfer \\
    \hdashline           
    T1    & Warm Transfer  & Warm Transfer  & Warm Transfer & Warm Transfer \\
    T2    & Warm Transfer & Cold Transfer & Warm Transfer & Warm Transfer \\
    T3w        & Cold Transfer & Cold Transfer & Warm Transfer & Warm Transfer\\
    T3c    & Warm Transfer & Cold Transfer & Warm Transfer & Cold Transfer\\
    T4   & Cold Transfer & Cold Transfer & Warm Transfer & Cold Transfer \\    
    T5   & Cold Transfer & Cold Transfer & Cold Transfer & Cold Transfer\tikzmark{end} \\
    \bottomrule
    \end{tabular}%
\begin{tikzpicture}[overlay,remember picture]
 \draw[<->,dotted] let \p1=(start), \p2=(end) in ($(\x1-116mm,\y1)+(0.3,0.2)+(0,-1.5mm)$) -- node[label=left:lowest service quality,align=center,yshift=-15mm,xshift=0mm ] {} node[label=left:highest service quality,align=center,yshift=15.5mm,,xshift=0mm ] {} ($(\x1-116mm,\y2)+(0.3,0.2)+(0,-1.5mm)$);
  \end{tikzpicture}
}
\end{table}%

\subsection{Terminal Conditions and Stationarity} \label{subsec: terminal conditions}
While there are many ways to specify terminal conditions, we tailor them such that the policy is stationary over $T$, making the optimal policy independent of the amount of time left in the agent's shift. This allows us to characterize both finite-horizon shifts and 24/7 operations using the same policy. In particular, the dynamic program (agent shift) terminates at the last decision epoch before taking the optimal action would otherwise result in $t$ exceeding $T$. If the agent is idle at this last decision epoch, the shift terminates at $T$ and the oncoming agent proceeds at $T+1$. If the agent has just finished an unsuccessful attempt at or before $T$, then the agent hands off the request to the oncoming agent, who starts the new shift and continues to handle the request optimally. Thus, as we show below, we specify the conditions such that the optimal actions of the agents are consistent throughout the shifts.

Let $G(X,Q,A)$ be a constant used in defining the value function under the optimal stationary policy at state $(X,Q,A)$. Since there are four congestion states for $X \in \{1,2,\cdots,S\}$ and we effectively collapse the four congestion states into two each for $X=0^i$ in (2.1) and for $X=0^w$ in (2.4), there are $4S + 2 + 2 = 4S+4 = 4(S+1)$ constants to be specified, which we refer to as the terminal conditions. Also, let $R^*$ be the revenue (net of agent and expert costs) per unit time under the optimal stationary policy. Then we define the solution to the dynamic program as \begin{equation} \label{eq:Terminal} V_t(X,Q,A) = R^*(T+1-t) + G(X,Q,A)\end{equation}

The term $R^* (T+1-t)$ is the pro-rated payment for the time remaining in the program, and the constant captures the expected incremental value of being in a given congestion state $(Q,A)$, given the agent's current status ($X$). In the proof of Theorem 2 below we specify the conditions necessary to induce stationarity, and refer to them as \textit{Type F} conditions, since they are affine functions of $R^*$. This leads us to the following:

\vspace{.1in}

\noindent \textbf{THEOREM 2} \textit{Under terminal conditions of Type F the optimal resolution policy is stationary, i.e., the optimal decision rule  after a given failed attempt does not depend on $t$.}

\vspace{.1in}

The economic meaning of these terminal conditions is as follows. For a given $t$, the value of (\ref{eq:Terminal}) increases in the number of failed attempts. This is because the request is getting closer to successful resolution. Moreover, for each $X$, $G(X,Q,A)$ is lowest for ($Q=0,A=0$) and highest for ($Q=1,A=1)$. This is because there is potential value in the presence of an arrival (since the agent may immediately begin processing a request) and potential value in an expert being available (since the agent may immediately begin performing a warm transfer). Also, the difference between $G(0^i,1,-)$ and $G(0^i,0,-)$ is $R^*/q$, the incremental expected opportunity cost of waiting for the next customer to arrive.  Analogously, the difference between $G(0^w,-,1)$ and $G(0^w,-,0)$ is $R^*/a$, the incremental expected opportunity cost of waiting for the expert to become available. This allows us to price out the incremental value of congestion, leading to the structural results next.



\subsection{Optimal Policy Under Stationarity} \label{subsec:stationarity}

Here we use the terminal conditions formulated in the proof of Theorem 2 to reduce the number of admissible decision rules for a given problem instance from 11 down to 3 or 4. The key is identifying when a cold transfer is preferred to a warm transfer for a given $(Q,A)$.  As in the proof of Theorem 3, we proceed by substituting the terminal conditions in (4) into the transfer arguments of the max operator in (2.5). Then, a cold transfer is preferred to a warm transfer for a given $(Q,A)$ if $$ G(0^i,Q,-) - G(0^w,-,A) \geq c^c - c^w.$$ Furthermore, substituting in $G(0^i,Q,-)$ and $G(0^w,-,A)$ from the proof of Theorem 2 yields $$R^*[\tau_w - 1 + (1-A)/a + Q/q] \geq c^c - c^w. $$ Adding and subtracting $1/q$ to the left-hand side and simplifying yields $$R^*[\tau_w + (1-q)/q + (1-A)/a] - R^*[(1-Q)/q] \geq c^c - c^w.$$

In the inequalities above, the right-hand side is the incremental savings from performing a warm transfer over a cold transfer (recall that $c^w < c^c$ as the expert requires less time to resolve a warm transfer). For a cold transfer to be preferred, these savings must exceed the foregone opportunity cost that a warm transfer induces relative to a cold transfer. This is given by the profit-adjusted difference between the first and second terms in brackets on the left-hand side. The first term in brackets is the expected time before the agent will recommence service after a warm transfer, and the second term is the analogous expected time after a cold transfer. This intuitive condition can be rearranged to show that a warm transfer is preferred to a cold transfer when the optimal profit $R^*$ is greater than $$R^* \geq (c^c - c^w)/[\tau_w - 1 + (1-A)/a + Q/a] \equiv \mathcal{R}(Q,A). $$ These intuitive insights are formalized in Theorem 3 below.

\vspace{.1in}

\noindent \textbf{THEOREM 3a} \textit{For a given $(Q,A)$ and any $X$, there exists a positive constant $\mathcal{R}(Q,A)$ such that when $R^*<\mathcal{R}(Q,A)$, a warm transfer is preferred to a cold transfer; otherwise, $R^*\geq \mathcal{R}(Q,A)$, and a cold transfer is preferred to a warm transfer.}

\vspace{.1in}

\noindent \textbf{THEOREM 3b} \textit{$R^*$ is such that $0 \leq \mathcal{R}(1,0) \leq \mathcal{R}(1,1) \leq \mathcal{R}(0,0) \leq \mathcal{R}(0,1)$, if $1/q \leq 1/a$; otherwise, $0 \leq \mathcal{R}(1,0) \leq \mathcal{R}(0,0) \leq \mathcal{R}(1,1) \leq \mathcal{R}(0,1)$.}

\vspace{.1in}

\noindent \textbf{THEOREM 3c} \textit{For a given problem instance, in an optimal resolution policy there are only up to four admissible decision rules: the continue rule, one or two of the four hybrid rules, and one of the six transfer rules.}

\vspace{.1in}

\noindent An implication of Theorem 3a is that for each congestion state $(Q,A)$, if the problem is not resolved at attempt $X$, the agent has two actions: continue or transfer to an expert; if transferring, then for every $X$, the transfer mode (cold transfer or warm transfer) is the same. In this sense, the choice is binary, but which of these actions is optimal may depend on $(Q,A)$. However, Theorem 3a is sufficient to provide further structure since the constants $\mathcal{R}(Q,A)$ can be ordered as in Theorem 3b; this ordering depends on the relative values of $q$ and $a$. Thus, for a given problem instance, there are five cases defined by ranges where $R^*$ may lie relative to the four $\mathcal{R}(Q,A)$ constants. Each range has a different set of preferred (more economical) transfer methods across the congestion states $(Q,A)$. Finally, Theorem 3c follows by determining, for each range, which of the 11 decision rules derived from Theorem 1 (listed in Table \ref{tab:decision_rules}) are potentially optimal under stationarity by admitting only the rules where, for each $(Q,A)$, the optimal action is either continue or to transfer the customer under the method that is more economical in that range. To elaborate on Theorem 3, we include Table \ref{tab:Ranges_app}. For ease of exposition, we only present the case where $1/q > 1/a$.

Table 2a provides the ranges, with the value of $R^*$ increasing from left to right. Table 2b lists the preferred transfer method for each congestion state. For example, in Case 1, $R^*$ is sufficiently low such that it is more economical to warm transfer the customer, regardless of congestion state. However, in Case 2, $R^*$ is sufficiently high such that it is more economical to cold transfer if there is an arrival and there is no expert available ($Q=1,A=0$), but sufficiently low such that it is more economical to warm transfer in the remaining congestion states. More generally, when $R^*$ is low, the revenue losses due to the inability to serve new arrivals are low. However, as $R^*$ increases, these revenue losses increase relative to the costs of cold transfers, resulting in cold transferring becoming increasingly preferred. At the extreme, in Case 5 it is more economical to cold transfer in all states.

\begin{table}[bt!] 
\caption{Ranges, Preferred Transfer Method, and Admissible Decision Rules Over Cases of $R^*$ ($1/q > 1/a$) \label{tab:Ranges_app}}
  \centering\footnotesize
    \begin{tabular}{ccccccc}
   \multicolumn{7}{c}{\textbf{Table 2a: Ranges}} \\ 
   \toprule
      \multicolumn{2}{c}{Case} & Case 1 & Case 2 & Case 3w & Case 4  & Case 5 \\
     \midrule
  \multicolumn{2}{c}{$R^*$} & $R^*<\mathcal{R}(1,0)$ & $\mathcal{R}(1,0)\leq R^* < \mathcal{R}(1,1)$ & $\mathcal{R}(1,1) \leq R^*<\mathcal{R}(0,0)$ & $\mathcal{R}(0,0)\leq R^* <\mathcal{R}(0,1)$ & $\mathcal{R}(0,1) \leq R^*$ \\ 
     \bottomrule
   \\  \multicolumn{7}{c}{\textbf{Table 2b: Preferred Transfer Method}} \\
      \toprule
    $Q$ & $A$  & Case 1 & Case 2 & Case 3w & Case 4 & Case 5 \\
    \midrule
    0 & 0 & Warm & Warm & Cold & Cold & Cold \\
    1 & 0 & Warm & Cold & Cold & Cold & Cold \\
    0 & 1 & Warm & Warm & Warm & Warm & Cold \\
    1 & 1 & Warm & Warm & Warm & Cold & Cold \\
     \midrule
    \\  \multicolumn{7}{c}{\textbf{Table 2c: Admissible Decision Rules}} \\
       \toprule
     Label & Rule & Case 1 & Case 2 & Case 3w & Case 4 & Case 5 \\
      \midrule 
     C & [n\scalebox{0.36}[1]{ },\scalebox{0.36}[1]{ }n\scalebox{0.36}[1]{ },\scalebox{0.36}[1]{ }n\scalebox{0.36}[1]{ },\scalebox{0.36}[1]{ }n] & \checkmark &  \checkmark & \checkmark & \checkmark & \checkmark \\

     \hdashline
     H1 & [n\scalebox{0.20}[1]{ },\scalebox{0.20}[1]{ }n\scalebox{0.20}[1]{ },\scalebox{0.20}[1]{ }w\scalebox{0.20}[1]{ },\scalebox{0.20}[1]{ }w] & \checkmark  & \checkmark & \checkmark & & \\
     H2 & [n\scalebox{0.23}[1]{ },\scalebox{0.23}[1]{ }c\scalebox{0.23}[1]{ },\scalebox{0.23}[1]{ }w\scalebox{0.23}[1]{ },\scalebox{0.23}[1]{ }w] & & \checkmark & \checkmark & & \\
     H3 & [n\scalebox{0.37}[1]{ },\scalebox{0.37}[1]{ }c\scalebox{0.37}[1]{ },\scalebox{0.37}[1]{ }w\scalebox{0.37}[1]{ },\scalebox{0.37}[1]{ }c] & & & & \checkmark & \\     
     H4 & [n\scalebox{0.44}[1]{ },\scalebox{0.44}[1]{ }c\scalebox{0.44}[1]{ },\scalebox{0.44}[1]{ }n\scalebox{0.44}[1]{ },\scalebox{0.44}[1]{ }c] & & & & \checkmark & \checkmark \\

     \hdashline
     T1 & [w,w,w,w] & \checkmark & & & & \\
     T2 & [w\scalebox{0.13}[1]{ },\scalebox{0.13}[1]{ }c\scalebox{0.13}[1]{ },\scalebox{0.13}[1]{ }w\scalebox{0.13}[1]{ },\scalebox{0.13}[1]{ }w] & & \checkmark & & &  \\
     \scalebox{2.12}[1]{ }T3w & [c\scalebox{0.27}[1]{ },\scalebox{0.27}[1]{ }c\scalebox{0.27}[1]{ },\scalebox{0.27}[1]{ }w\scalebox{0.27}[1]{ },\scalebox{0.27}[1]{ }w] & & &\checkmark & &  \\
     T4 & [c\scalebox{0.40}[1]{ },\scalebox{0.40}[1]{ }c\scalebox{0.40}[1]{ },\scalebox{0.40}[1]{ }w\scalebox{0.40}[1]{ },\scalebox{0.40}[1]{ }c] & & & & \checkmark &  \\
     T5 & [c\scalebox{0.53}[1]{ },\scalebox{0.53}[1]{ }c\scalebox{0.53}[1]{ },\scalebox{0.53}[1]{ }c\scalebox{0.53}[1]{ },\scalebox{0.53}[1]{ }c] & & & & & \checkmark \\
     \bottomrule
    \end{tabular}%
\begin{tablenotes}
\item n = continue, w = warm transfer, c = cold transfer
\end{tablenotes}
\end{table}%

Finally, given the preferred transfer type, we can construct Table 2c, where we indicate, by range, which decision rules are admissible. We highlight the two corner cases. Case 1 contains problem scenarios where the firm's optimal resolution policy is to provide the most personalized service possible. So, after a given failure, the admissible transfer rule is T1, in which the agent must perform a warm transfer, regardless of $Q$ and $A$, and the admissible hybrid rule is H1, wherein the agent performs a warm transfer if there is expert availability ($A = 1$). At the other extreme is Case 5, where the firm provides the least-personalized service, concentrating on throughput instead. In this case, the only admissible transfer rule is T5, wherein the agent must cold transfer the customer, regardless of $Q$ and $A$, and the only admissible hybrid rule is H4, wherein the agent cold transfers contingent on the presence of an arrival ($Q = 1$). One consequence of Theorem 3 is that in any problem instance the number of admissible policies is limited to three or four.  While insightful, this structural result leaves issues of fast running times unresolved.  As discussed in the next section, to develop fast algorithms, we propose using a threshold policy as a heuristic. 

\section{Heuristics} 
In this section we extend our modeling in two ways. We first evaluate system performance under an intuitive threshold policy that further reduces policy complexity; we find that the heuristic approach performs quite well and is optimal under a sufficiency condition which has a shortest processing time interpretation. We then extend our analysis to include a waiting room whose operationalization we show under a two-attempt model without the warm transfer option. This allows us to draw additional insights into policy performance under queueing dynamics. 

\subsection{Threshold Policy Heuristic} While our structural analysis has reduced the optimal policy search to a manageable number of decision rules, the number of admissible policies still grows exponentially in $S$. As a polynomial-time alternative to the optimal algorithm, we examined performance using a threshold policy.  A policy has a threshold structure when, for a given $(Q,A)$, if the policy is to transfer after failed attempt $X$, then it would also be to transfer for all subsequent failed attempts. Intuitively, under a threshold policy, the decision-making sequence for a given request is: for a given congestion state, continue until a specified number of attempts; thereafter, perform the transfer type specified for that congestion state. From a practical standpoint, this would allow for easy management communication of the policy structure and eliminate the burden of retaining a decision rule for each attempt/congestion-state combination.

In EC.2 we measure the performance of threshold policies as a heuristic solution for 1,200,000 randomly generated problem instances, each for $S = 3$, $S = 4$, $S = 5$, and $S = 6$.  Table \ref{tab:Heuristic Performance} summarizes the relevant performance measures from our analysis. 

\begin{table}[h!]\footnotesize
\caption{Performance Measures of Threshold Policy Heuristic\label{tab:Heuristic Performance}}
   \centering\footnotesize 
   \begin{tabular}{l|cccc}
    \toprule
          & \multicolumn{4}{c}{Number of Potential Solutions ($S$)} \\
    Performance Measure & $S = 3$ & $S = 4$ & $S = 5$ & $S=6$ \\
    \midrule
    \% Instances Threshold Optimal & 97.46\% & 96.58\% & 96.02\% & 95.62\% \\
    Average Optimality Gap & 0.005\% & 0.006\% & 0.006\% & 0.005\% \\
    Maximum Optimality Gap & 1.888\% & 1.818\% & 1.634\% & 1.575\% \\
    \bottomrule
    \end{tabular}%
  {}
\end{table}%

In over 95\% of the instances, a threshold policy is optimal, with an average optimality gap of no greater than 0.006\%.  Notably, while the fraction of instances  where a threshold policy is optimal is decreasing in $S$, the optimality gap remains stable; this is because the optimality gap in instances where a threshold policy was not optimal is decreasing in $S$. Additionally, worst-case performance measured by the maximum optimality gap decreases in $S$, suggesting that the threshold heuristic performance improves as the computational burden for finding the optimal policy increases. Overall, this analysis suggests that threshold policies achieve optimal or close-to-optimal performance in the majority of scenarios. 

Since we find that threshold policies often achieve optimal or close-to-optimal performance, we delved deeper into the policy structure to explore when threshold policies are provably optimal. Because processing times, adjusted for uncertainty, play a prominent role in the transfer decision we studied many variants of the weighted shortest process time (WSPT) rule as explicated in Chapter 9 of Pinedo (1995). Ultimately, we found an interpretable variant given next.

\vspace{.1in}

\noindent \textbf{THEOREM 4} \textit{A threshold policy is an optimal solution if either of the following conditions holds:}

\vspace{.1in}

\noindent \textit{(i) Condition C-WSPT-a: If $1/a > 1/q$ and for $X = 2, 3, \cdots, S,$ \\ $\tau_X/\rho_X - \tau_{X-1}/\rho_{X-1} \leq  (1/a)/\rho_X + [(1-a)/a](1- \rho_{X-1})/\rho_{X-1}$; or}

\vspace{.1in}

\noindent \textit{(ii) Condition C-WSPT-q: If $1/a \leq 1/q$ and  for $X = 2, 3, \cdots, S,$ \\ $\tau_X/\rho_X - \tau_{X-1}/\rho_{X-1} \leq  (1/q)/\rho_X + [(1-q)/q](1- \rho_{X-1})/\rho_{X-1}.$}

\vspace{.1in}

Note that this variant of WSPT is appealing in that it gives a sufficient condition independent of all economic parameters of the problem.  Interestingly, the weights depend on uncertainty but via conditional probabilities, which we interpret as a conditional risk-adjusted WSPT. See EC.2 for the proof of the above theorem and a further discussion of its intuition and connection to WSPT rules.

\subsection{Finite Waiting Room} We have thus far examined a system where arriving customers are not admitted if the agent is unavailable. However, given the prevalence of queueing in real-world service systems (in the form of finite and infinite waiting rooms), a brief examination of how its inclusion would impact our analysis is warranted. The addition of a waiting room to our model would allow us to capture these dynamics, but would also substantially increase the number of admissible policies as one would need to condition the transfer decision at a given attempt on the current queue length. While our approach would work in general, to keep focus we illustrate our approach using a waiting room of capacity $N$ where only cold transfers are permitted and at most two resolution attempts are required.

To simplify analysis, we examine the problem under an infinite horizon, which bypasses the need to induce policy stationarity through terminal conditions. Given this setup, the following policies are admissible: Never Transfer, Always Transfer and Transfer when $Q\geq n, n = 1, 2, \cdots, N$ ($Q$ is now the number of customers in the waiting room plus up to one current arrival). Under an infinite horizon, each of these policies can be characterized as a Discrete Time Markov Chain (DTMC), yielding steady-state probabilities with which we can calculate the profit per period. We demonstrate how to perform these calculations in EC.3. Note that choosing $N=0$ yields the model presented in \textsection 4.

\begin{figure}[bt]
\begin{center}
	\caption{Profit Per Period by Arrival Intensity Under Admissible Policies for Waiting Room Capacity Six}\label{fig:Queueing_1}
	\includegraphics[trim = 0in 0in 0in 0in, clip, width=5.5in]{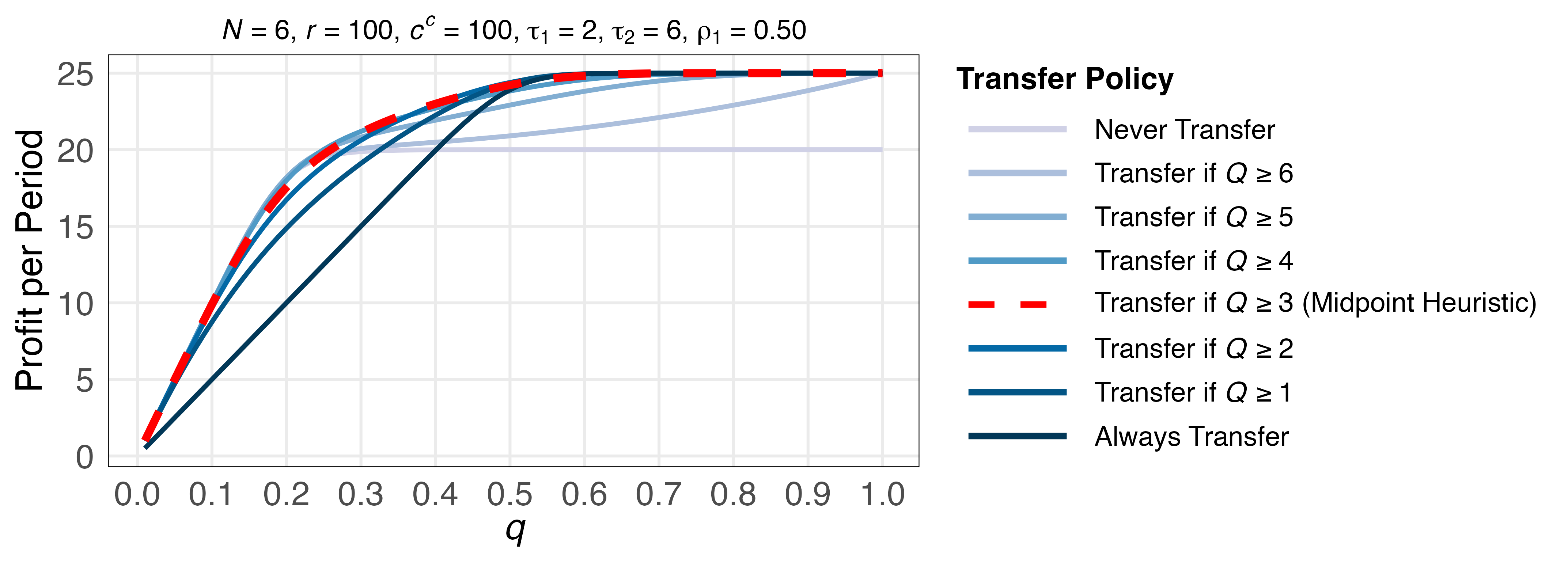}
\end{center}
\end{figure}

Using the DTMC structure, we perform numerical experiments with a waiting room size of six ($N=6$) to allow for sufficient demand buildup to illustrate the queueing dynamics. Consider first Figure \ref{fig:Queueing_1}. The figure shows the performance of all admissible policies, with the parameters reported above the plot area. The arrival intensity ($q$) is varied between 0.01 and 1.00 on the horizontal axis and profit per period is shown on the vertical axis.  For $q\in[0.01, 0.40]$, ``Never Transfer'' dominates ``Always Transfer''. In this range, arrival intensity is sufficiently low such that the cost savings from not transferring outweighs the extra revenue loss caused by the additional blocking resulting from longer service times and, therefore, higher utilization. Conversely, for $q\in[0.41, 1.00]$, the additional revenue from accepting new customers from the waiting room outweighs transfer costs, and ``Always Transfer'' dominates ``Never Transfer''.  Also note that for $q\in[0.02, 0.66]$, state-contingent policies are optimal. Moreover, among all policies averaged over $q$, the best is to transfer if $Q\geq 3$. This has an intuitive interpretation as the ``transfer when waiting room is half full'' policy, which allows for sufficient storage of demand but mitigates excessive customer blocking. Finally, note that each policy approaches an asymptotic profit per period based on its maximum service rate.\footnote{The state-contingent policies approach the same asymptotic profit as that of the ``Always Transfer'' policy since, under high arrival intensity, the waiting room is nearly always full, resulting in nearly every request being transferred if the first attempt fails. In other words, the state-contingent policies behave nearly the same as the ``Always Transfer'' policy under high intensity.}

\begin{figure}[tb]
\begin{center}
	\caption{Optimal Profit Per Period by Arrival Intensity with and without Waiting Room}\label{fig:Queueing_2}
	\includegraphics[trim = 0in 0in 0in 0in, clip, width=5.5in]{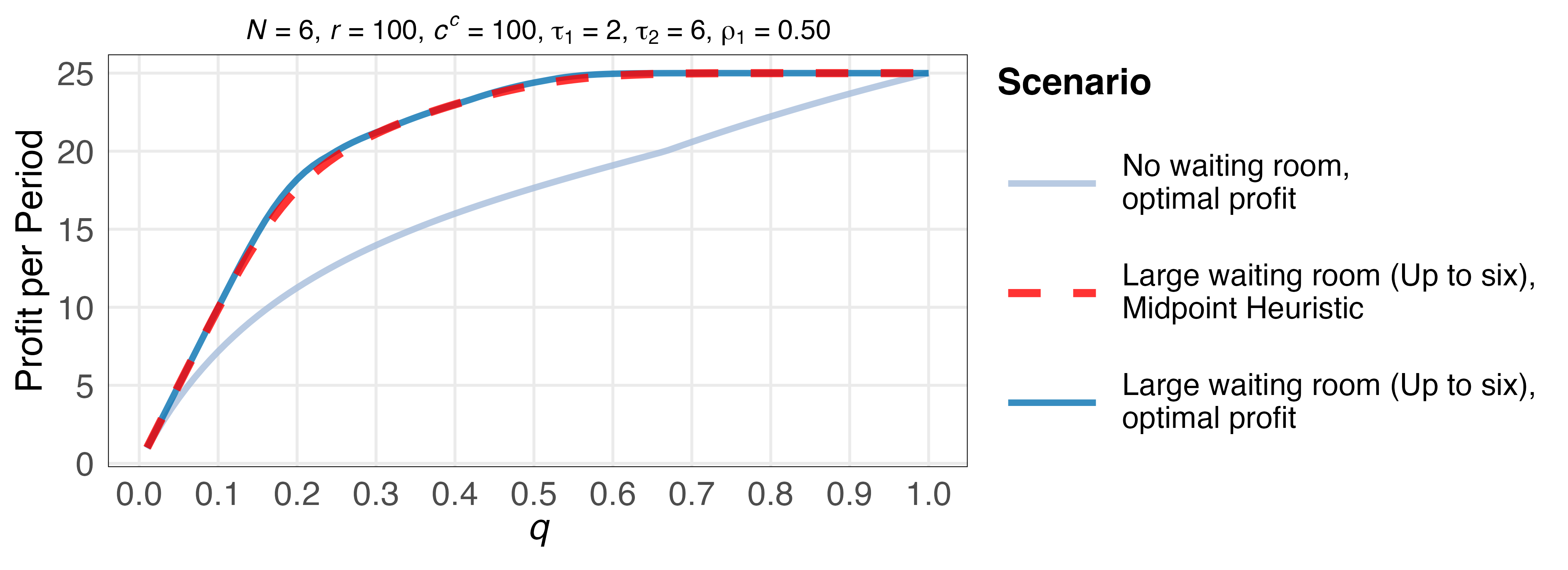}
\end{center}
\end{figure}

Figure \ref{fig:Queueing_2} shows how the presence of a waiting room affects profits. Specifically, we plot the optimal profit under three scenarios: 1) optimal profit of all three admissible policies with no waiting room, 2) profit with the waiting room under the Transfer when $Q\geq 3$ Policy, 3) optimal profit with the waiting room under all eight admissible policies (which we examined in Figure \ref{fig:Queueing_1}). Unsurprisingly, the inclusion of the waiting room increases profit for all values of $q$ as some demand may now be postponed rather than lost. However, the increase is minimal at the extremes as the waiting room is seldom used under very low intensity and nearly always full under high intensity. When comparing optimal profits in the waiting room scenarios, we see that restricting the policy search to ``Transfer when $Q\geq 3$'' results in nearly the same profit as the full set of admissible policies. This is good news for managers as it indicates that intuitive search heuristics could be quite effective when the policy space is large and customers are patient. 

\section{Application to AI-enabled Customer Service Design} 
In this section we turn to the firm's broader customer service design problem, which includes the strategic selection of service channels, the tactical choice of live-agent (gatekeeper) staffing levels and automated/AI capability investment, and the operational control of resolution policies. To maintain focus, we limit our analysis to a two-channel system with homogeneous customers, where the firm may offer a live-agent channel, an AI chatbot channel, or both. While the presented approach can be easily generalized to accommodate more than two channels or heterogeneous customers, we omit these extensions for the sake of brevity and clarity. As in the previous sections, we prioritize intuition and key results, with technical details provided in EC.4 - EC.6.

\subsection{Overview of Firm's Problem}
To examine the firm's service design problem, we revisit the profit-maximization problem from (\ref{eq:architecture}) in \textsection 2.1: \begin{align}\label{eq:architecture_2}
\max\limits_{k^{agent},\boldsymbol{D^{agent}},p^{bot}_{succ}} \pi^{agent}(k^{agent},\boldsymbol{D^{agent}},\lambda^{agent}(k^{agent},\boldsymbol{D^{agent}},p^{bot}_{succ})) + \pi^{bot}(p^{bot}_{succ},\lambda^{bot}(k^{agent},\boldsymbol{D^{agent}},p^{bot}_{succ})).
\end{align}

\noindent We have already shown in \textsection 3 - \textsection 4 how to optimally set the live-agent resolution policies, $\boldsymbol{D^{agent}}$.  In \textsection 6.2  we extend our model of request resolution to a multi-server setting, which will allow us to determine the optimal staffing levels, $k^{agent}$. In \textsection 6.3 we show how the firm can optimally set the chatbot performance level, $p^{bot}_{succ}$. Also note that because the problem in (\ref{eq:architecture_2}) is formulated from the firm's profit-maximization perspective, it must balance both cost and quality considerations. This is because $k^{agent}$, $\boldsymbol{D^{agent}}$, and $p^{bot}_{succ}$ give rise to channel-specific performance measures (such as transfer probabilities and expected service times), which in turn affect service quality.  We characterize these effects in \textsection 6.4, where we solve the customers' channel choice problem and calculate demand rates for each channel $(\lambda^{agent}, \lambda^{bot})$ as a response to exogenously given service performance measures. Then, in \textsection 6.5 we align the firm's and the customers' problems by endogenizing the arrival rates to each channel. To do so, we set the demand rates expected by the firm equal to the arrival rates resultant from aggregate customer decisions (setting $\lambda^{agent} = q^{agent}/k^{agent}$, $\lambda^{bot} = q^{bot}$) and solve (\ref{eq:architecture_2}) by evaluating it over all channel mix combinations, staffing levels, chatbot capabilities, and optimized resolution policies within each channel. Finally, in \textsection 6.6 we perform numerical experiments over a range of plausible bot training and wage cost structures to gain managerial insights.

\subsection{Multiple Servers ($k^{agent}>1$)} To this point, our analysis has focused on the optimal control problem of a single live agent under exogenous demand. In practice, most live customer service channels employ multiple agents to handle demand, which introduces an additional layer of complexity and the decision variable of agent staffing levels. Increasing the number of agents could enable the firm to deliver higher service quality by changing the resolution policy such that agents make additional resolution attempts before transferring (or warm transfer rather than cold transfer after failure). Moreover, the joint staffing and resolution policy problem also interacts with the channel selection problem as the inclusion of automated channels such as AI chatbots could divert demand away from the live channels and reduce (or fully eliminate) the need for live agents.

Extending our analysis to a multiserver setting would require us to condition the focal agent's optimal action on the status of all other agents, rendering practical analysis intractable; for example, the asymptotic $M/M/1$ approach based on \cite{abate1995exponential,abate1996exponential} cannot be used here because our controls are state-dependent. We thus propose an alternative routing scheme that maintains the same admissible resolution policies as the single agent model. In particular, customers who enter the channel are randomly assigned to one of $k^{agent}$ agents (irrespective of status), thereby retaining for each agent the Bernoulli arrival process described in \textsection 3.1. This is akin to splitting traffic into dedicated queues and assigning work using a round-robin system. Once incoming customer traffic is allocated, each agent channel structure is that of the model of Sections 3 and 4, allowing us to focus on different customer service designs, and the inherent trade-offs between offering different channel combinations, chatbot training, staffing, and resolution policies.

\subsection{The Chatbot Channel} Our interviews with BlackBeltHelp (described in \textsection 1) suggest that AI developers program chatbots to make a limited number of attempts to diagnose and resolve each request before transferring the customer to a human expert. This is similar to the $S$-attempt approach that we used to model the live-agent channel (\textsection 3). In the case of the chatbot channel, the developer conducts a cost-benefit analysis to choose which of the $S$ attempts to program the chatbot to make, and to what degree of reliability. Since the chatbot's opportunity cost of time is negligible, the chatbot continues to make attempts until all attempts for a given request are exhausted, before transferring the customer. As a result, the chatbot's resolution policy is fully characterized by its resolution probability, $p^{bot}_{succ}$, where the development cost function is increasing and convex in $p^{bot}_{succ}$. A convex cost is directly supported by the extensive computer science literature discussing the diminishing returns of training domain-specific chatbots \citep[][and references there]{hoffmann2022}. We choose the form $c^{bot}(p^{bot}_{succ}) = a^{bot}(100\cdot p^{bot}_{succ})^{b^{bot}}$, which gives us flexibility to examine scenarios with different base costs  (scale parameter $a^{bot}$) and different rates of increase (exponent $b^{bot}$).

The chatbot's resolution policy is further simplified by the limited nature of its action set (relative to that of the live agent). In particular, if the chatbot fails to resolve the request, the customer \textit{must} be cold transferred to the expert. This is because the chatbot cannot perform warm transfers.  Given that the chatbot will make up to all programmed attempts before transferring to the expert, it is as if the chatbot is programmed as a gatekeeper whose optimal policy is static in that it performs the sequential request resolution procedure (from \textsection 3) and then transfers upon failure. Note that such a static protocol is a special instance of Case 5 of Theorem 3 in Table \ref{tab:Ranges_app}. Finally, from the customer's perspective, there is no waiting time to access the chatbot. The chatbot admits all customers that arrive and they are served immediately without waiting.

\subsection{Customer Channel Choice Model}
\begin{figure}[b]
\begin{center}
	\caption{Customer Channel Choice Model}\label{fig:Discrete Choice Main}
	\includegraphics[trim = 0in 0in 0in 0in, clip, width=6.0in]{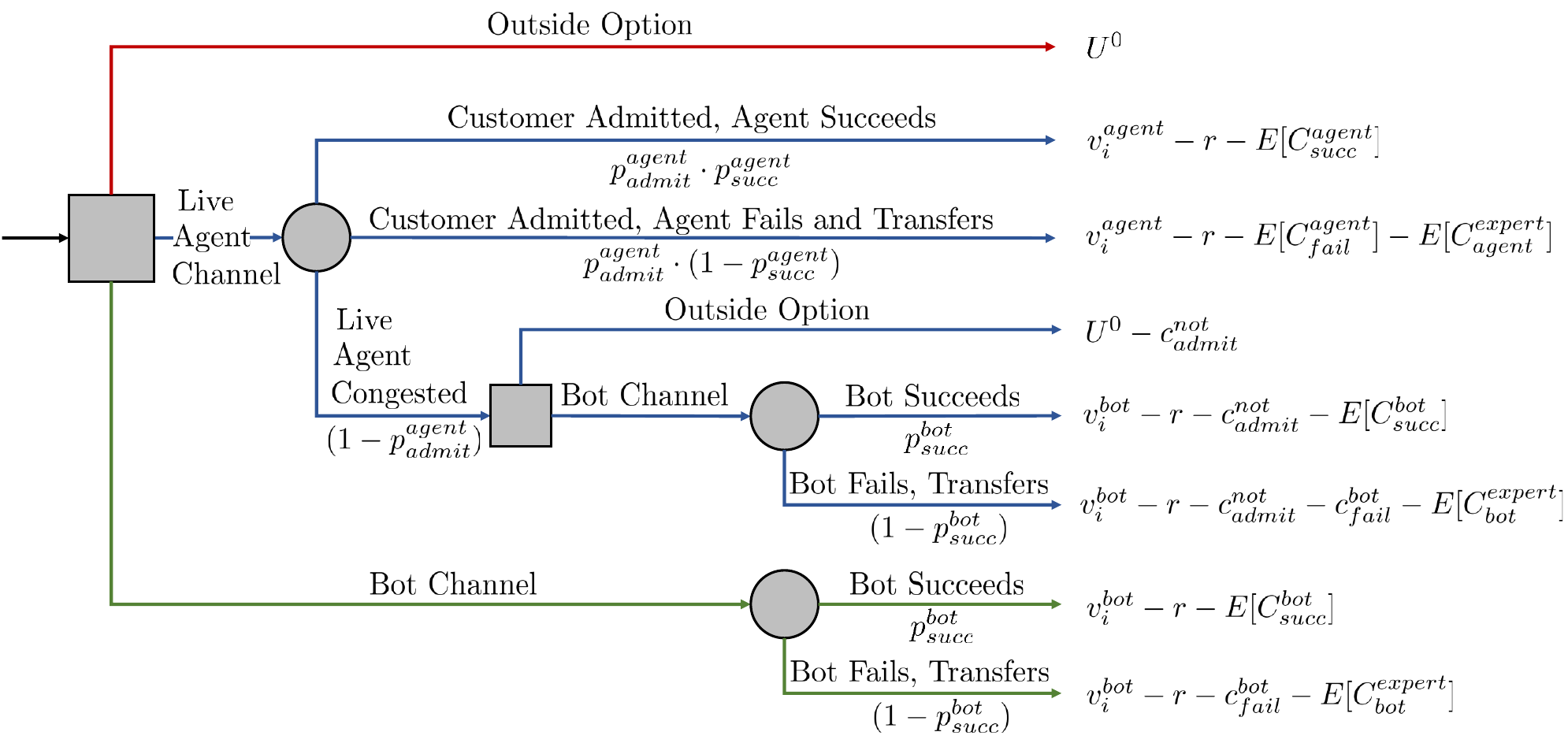}
\end{center}
\end{figure}

The staffing level, resolution policy, and chatbot capability ($k^{agent},\boldsymbol{D^{agent}},p^{bot}_{succ}$) give rise to channel performance measures; specifically, the customers' admission probability (in the live-agent channel),  transfer probabilities, and expected service times. Customers respond to these measures by choosing the channel (if any) that maximizes their utility. The resultant demand model is shown in Figure \ref{fig:Discrete Choice Main}.

Each customer $i$ chooses between joining the live-agent channel, the chatbot channel, and a fixed outside option $U^0$, knowing that once admitted, the customer is assured of request resolution. However, from the customer's perspective, the outcome and the amount of time spent in each channel are uncertain. We represent this uncertainty with the circles (random nodes) in Figure \ref{fig:Discrete Choice Main}. The live-agent channel has three possible outcomes: the agent that the customer is routed to succeeds, the agent transfers the customer to the expert, and the agent is busy serving another customer. In the latter case, the customer is left with the choice to leave the system or to contact the chatbot. (Consistent with \textsection 3, customers who are not admitted do not wait). The chatbot channel is analogous, with the difference being that it is never congested. Customers act on their idiosyncratic technology preferences (channel-specific service valuations  $v^{agent}_i$ and $v^{bot}_i$). Their responses also depend on potential aversion to transfers (included in the $E[C^{expert}]$ terms), the relative comparison of waiting costs ($E[C]$ terms), and possible disutility of not being admitted to the live-agent channel ($c^{not}_{admit}$). Note that the transfer aversion captured in $E[C^{expert}]$ is set higher for cold transfers than warm transfers as suggested from our customer survey in \textsection 1. In EC.4 we fully specify the utilities and solve for the demand rates for each channel, $\lambda^{agent}$ and $\lambda^{bot}$, as a function of performance measures communicated to customers.

\subsection{Customer Service Design Problem}
We have so far assumed that the firm's problem and the customers' problem are solved independently. That is, the firm's problem is solved under exogenously given arrival rates, while the channel choice model in \textsection 6.4 is solved under exogenously given performance measures. Linking the two problems requires that we endogenize channel demand such that we end up at a ``rational expectations'' equilibrium in which the firm delivers that which is promised to the customer, who in turn anticipates the fulfillment of this promise. We do this by solving for $\lambda^{agent} = q^{agent}/k^{agent}$ and $\lambda^{bot}=q^{bot}$ such that arrival rates from the resolution model and demand rates from the customer choice model are set equal under a given choice of resolution policies ($\boldsymbol{D^{agent}},p^{bot}_{succ})$. To build managerial intuition, we do this for the special case of uniformly distributed service valuations ($v^{agent}_i$, $v^{bot}_i$). See EC.5 for details. 

Once the profit maximization problem in (5) is fully specified, it can be solved by jointly evaluating up to four decisions: 1) the strategic decision of which gatekeeper channel(s) to deploy -- live-agent-only, chatbot-only, or both, 2) and 3) the tactical decisions of what staffing level to set in the live-agent channel ($k^{agent}$), and to what extent the chatbot is trained ($p^{bot}_{succ}$), and 4) the operational decision of which live-agent resolution policy to employ ($\boldsymbol{D^{agent}}$). In EC.6 we formulate the firm's profit function in (\ref{eq:architecture_2}) under the above inputs. In essence, for each staffing level (including $k^{agent}=0$), the firm iterates through each admissible live-agent resolution policy from Table \ref{tab:decision_rules} and each potential chatbot success probability value (including $p^{bot}_{succ} = 0$) to find the inputs that optimally balance channel demand, chatbot training costs, and expert resolution costs. Then the firm selects the most profitable combination.

\subsection{Numerical Illustration}

For the purposes of illustration, we set $S=2$ and evaluate 12 scenarios that differ in the shape of the training cost function for the bot, as well as the cost of labor (agent wage $c^{wage}$). The parameters used in the illustration are described in the notes of Table \ref{tab:Scenarios}.\footnote{In addition to the parameters described in the table notes of Table \ref{tab:Scenarios}, we assume that the firm pays the expert 3.5 per period to process requests. Chatbot processing times are fixed and exogenous at 6 periods. Customers incur a disutility of 1 for each period spent with the agent or bot, and a non-admission cost of 1 if not admitted to the live-agent channel ($c^{not}_{admit} = 1$). To reflect the lower perceived service quality of being transferred, the per period disutility increases to 1.25 after a warm transfer and to 2 after a cold transfer.}

The optimal customer service design, which includes optimal staffing level $k^{agent}$, optimal agent resolution policy $D^{agent}$ and optimal chatbot performance $p^{bot}_{succ}$, is summarized in Table \ref{tab:Scenarios}. Consider first Row (i), where the base cost is low and the cost increases at a slow rate. These scenarios are representative of a firm that can, with some effort, delegate the majority of customer problems to a chatbot. This can be reflective of a relatively large e-commerce company specializing in consumer electronics, where a sophisticated chatbot can handle order tracking, provide detailed product information, process return requests, and offer basic troubleshooting for common issues. In this scenario, the optimal solution is to train the chatbot up to a relatively high level, with $p^{bot}_{succ}$ ranging between 0.83 and 0.90 in the optimum. Further, as one may expect, the optimal chatbot investment increases as $c^{wage}$ increases. Indeed, in the case of medium and high labor costs, the optimal strategy is to have no human gatekeepers, i.e., $k^{agent}=0$. 

While Row (i) may be appealing from the firm's perspective, in the majority of real-life applications, technology has not yet achieved near-perfect capabilities, suggesting that Rows (ii)-(iv) may be more reflective of the current chatbot economics.\footnote{A recent survey in \cite{kagan2023} finds that customer service chatbots are currently able to resolve between 30 and 40 percent of customer inquiries in most consumer settings.} In these scenarios, the costs of providing a high $p^{bot}_{succ}$ can be prohibitively high, which may be reflective of services such as banking or healthcare, where complex and personalized interactions are often required. For example, in banking, chatbots might handle basic account inquiries or transaction histories, but struggle with nuanced financial advice or complex loan applications. In healthcare, chatbots may schedule appointments or provide general health information, but cannot replace the expertise of medical professionals for diagnosis or treatment plans. In these scenarios, we observe that the optimal $p^{bot}_{succ}$ reaches at most 0.59 (Scenario H(ii)). Indeed, if the chatbot development costs are sufficiently high and agent wages are low (Scenario L(iv)), an agent-only architecture with a large number of agents ($k^{agent}=7$) paired with the most customer-friendly policy of never transferring ($D^{agent}= C$) becomes optimal, with chatbots no longer being viable.

\begin{table}[h!]
\renewcommand{\arraystretch}{1.5}
\caption{Optimal Customer Service Design under Various Live-Agent and Chatbot Development Costs}
\label{tab:Scenarios}
\centering \footnotesize
\begin{tabular}{>{\raggedright\arraybackslash}L{5cm}|>{\centering\arraybackslash}p{3.5cm}|>{\centering\arraybackslash}p{3.5cm}|>{\centering\arraybackslash}p{3.5cm}}
\toprule
 & \multicolumn{3}{c}{\textbf{Live Gatekeeper Wage ($c^{wage}$)}} \\
\cmidrule{2-4}
\textbf{Bot Training Cost Scenario} &  \textbf{(L)ow} & \textbf{(M)edium} & \textbf{(H)igh} \\
\midrule
(i) Low base cost,  slowly increasing & 
$k^{agent} = 2$, $\boldsymbol{D^{agent}} = C$, $p^{bot}_{succ} = 0.83$ & 
$k^{agent} = 0$, $\boldsymbol{D^{agent}} = N/A$, $p^{bot}_{succ} = 0.90$ & 
$k^{agent} = 0$, $\boldsymbol{D^{agent}} = N/A$, $p^{bot}_{succ} = 0.90$ \\
(ii) High base cost, slowly increasing & 
$k^{agent} = 3$, $\boldsymbol{D^{agent}} = C$, $p^{bot}_{succ} = 0.52$ & 
$k^{agent} = 1$, $\boldsymbol{D^{agent}} = C$, $p^{bot}_{succ} = 0.56$ & 
$k^{agent} = 0$, $\boldsymbol{D^{agent}} = N/A$, $p^{bot}_{succ} = 0.59$ \\
(iii) Low base cost, rapidly increasing & 
$k^{agent} = 4$, $\boldsymbol{D^{agent}} = C$, $p^{bot}_{succ} = 0.28$ & 
$k^{agent} = 2$, $\boldsymbol{D^{agent}} = C$, $p^{bot}_{succ} = 0.29$ & 
$k^{agent} = 1$, $\boldsymbol{D^{agent}} = T5$, $p^{bot}_{succ} = 0.29$ \\
(iv) High base cost,   rapidly increasing & 
$k^{agent} = 7$, $\boldsymbol{D^{agent}} = C$, $p^{bot}_{succ} = 0.00$ & 
$k^{agent} = 2$, $\boldsymbol{D^{agent}} = C$, $p^{bot}_{succ} = 0.21$ & 
$k^{agent} = 1$, $\boldsymbol{D^{agent}} = T5$, $p^{bot}_{succ} = 0.21$ \\

\bottomrule
\end{tabular}
\vspace{1mm}

\begin{minipage}{1.00\textwidth}
\textit{Notes:} The bot training cost function is $a^{bot}(p^{bot}_{succ})^{b^{bot}}$ with parameters: (i) $a^{bot}=0.00005$, $b^{bot}=2.60$; (ii) $a^{bot}=0.00005$, $b^{bot}=3.00$; (iii) $a^{bot}=0.00009$, $b^{bot}=2.60$; (iv) $a^{bot}=0.00009$, $b^{bot}=3.00$, $c^{wage}\in\{0.5,0.9,1.3\}$, $r = 20$, $\tau_1 = 6$, $\tau_2=15$, $\tau_w = 3$, $\rho_1 = 0.7$, $\rho_2 = 1$, $a=0.8$, ($v^{agent}_i,v^{bot}_i) \sim U[0,100]$. 
\end{minipage}
\end{table}

\begin{figure}[h]
\begin{center}
	\caption{Numerical Illustration (Scenario M(iii))}\label{fig:Use Case}
	\includegraphics[trim = 0in 0in 0in 0in, clip, width=6.5in]{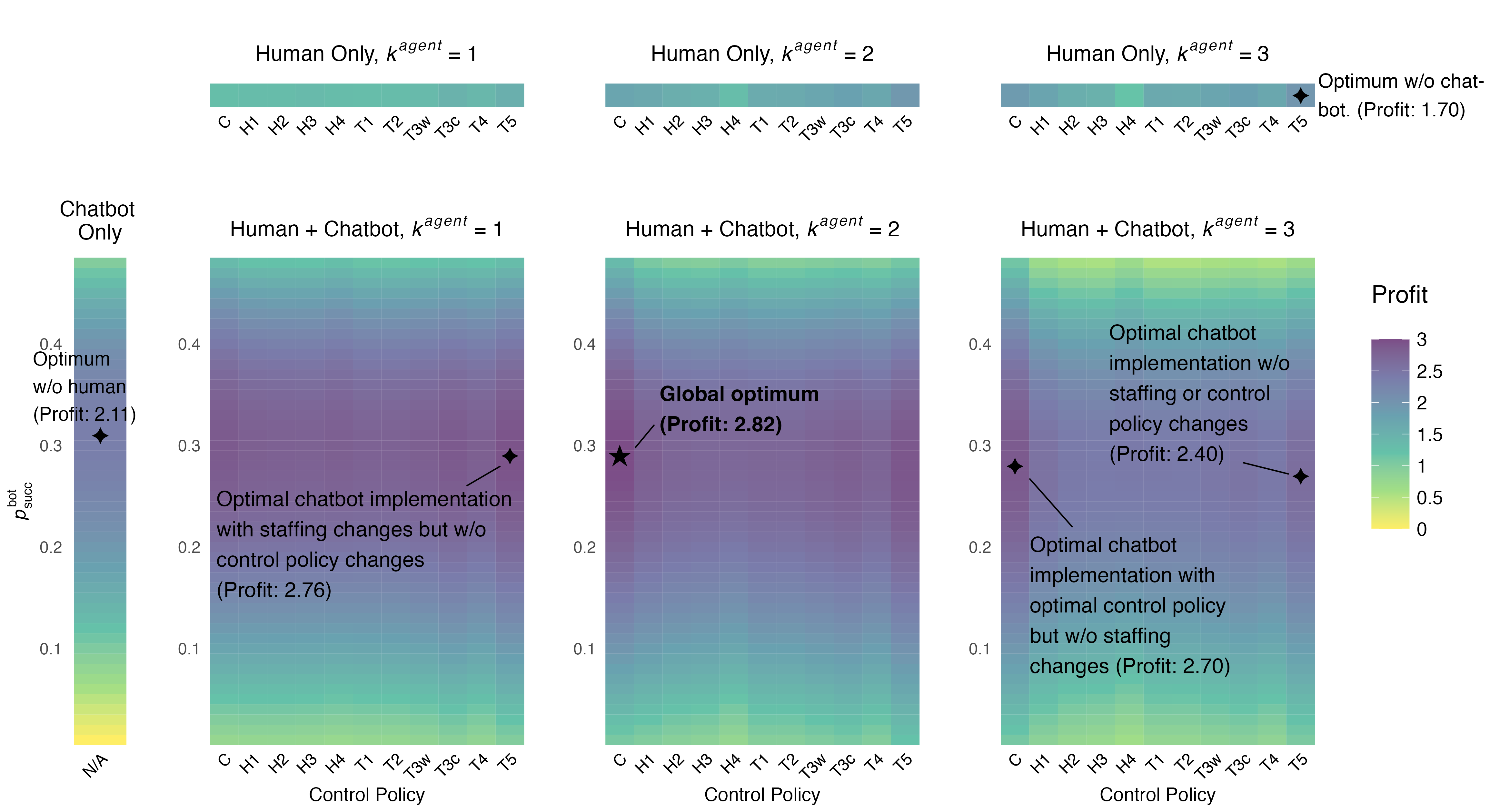}
\end{center}
\end{figure}

To gain deeper insights, we zoom into Scenario M(iii) from Table \ref{tab:Scenarios}. For this scenario, we illustrate in Figure \ref{fig:Use Case} the firm's profit per period for different combinations of $p^{succ}_{bot}$, $k^{agent}$, and $D^{agent}$. The top three heat bars show profits without chatbots for $k^{agent}={1,2,3}$. The leftmost (vertical) heat bar represents the chatbot-only scenario ($k^{agent}=0$). The three heat maps on the bottom right display profits with a hybrid, human+chatbot strategy and $k^{agent}={1,2,3}$.  

Without chatbot technology, the optimal staffing is $k^{agent}=3$ and the optimal resolution policy is T5 ("Always Transfer"), yielding a profit of 1.70. Introducing chatbots while maintaining the same staffing and resolution policy increases profits by 41.2\% to 2.40. Further improvements can be achieved by switching to resolution policy C ("Never Transfer"), increasing profits by an additional 17.6\% to 2.70. This is achieved through improved customer service quality, which in turn increases demand and revenue. Notably, the globally optimal design involves reducing the number of gatekeepers to two ($k^{agent}=2$) and adopting the customer-friendly resolution policy C. Moreover, across all scenarios that involve chatbot implementation, the optimal $p^{bot}_{succ}$ is remarkably stable at around 0.30, regardless of resolution policy and $k^{agent}$.

An important managerial takeaway from this numerical analysis is that any design involving both humans and chatbots can obtain near-optimal performance, particularly when the appropriate resolution policy is selected. Near-optimal performance can be achieved by a low staffing policy ($k^{agent}=1$) paired with a low service quality policy (``Always Transfer'', i.e., T5). Alternatively, good performance can also be achieved by a maximum staffing policy ($k^{agent}=3$) paired with a high service quality (``Never Transfer'', i.e., C). Hybrid channel architectures thus give firms the flexibility to position themselves as a cost or a service quality leader. In contrast, strategies that involve chatbots only or humans only are less flexible and lead to lower profits. More broadly, these comparisons highlight that firms need to consider both the strategic implications of AI development investments and the indirect effects on staffing and service delivery processes, in order to balance service quality and costs.

\section{Conclusion} \label{sec:Conclusion}
\begin{figure}[h]
\begin{center}
	\caption{Three Layers of Customer Service operations} \label{fig:strat:tact:oper} 
	\smallskip
	\includegraphics[trim = 0in 0in 0in 0in, clip, width=4.50in]{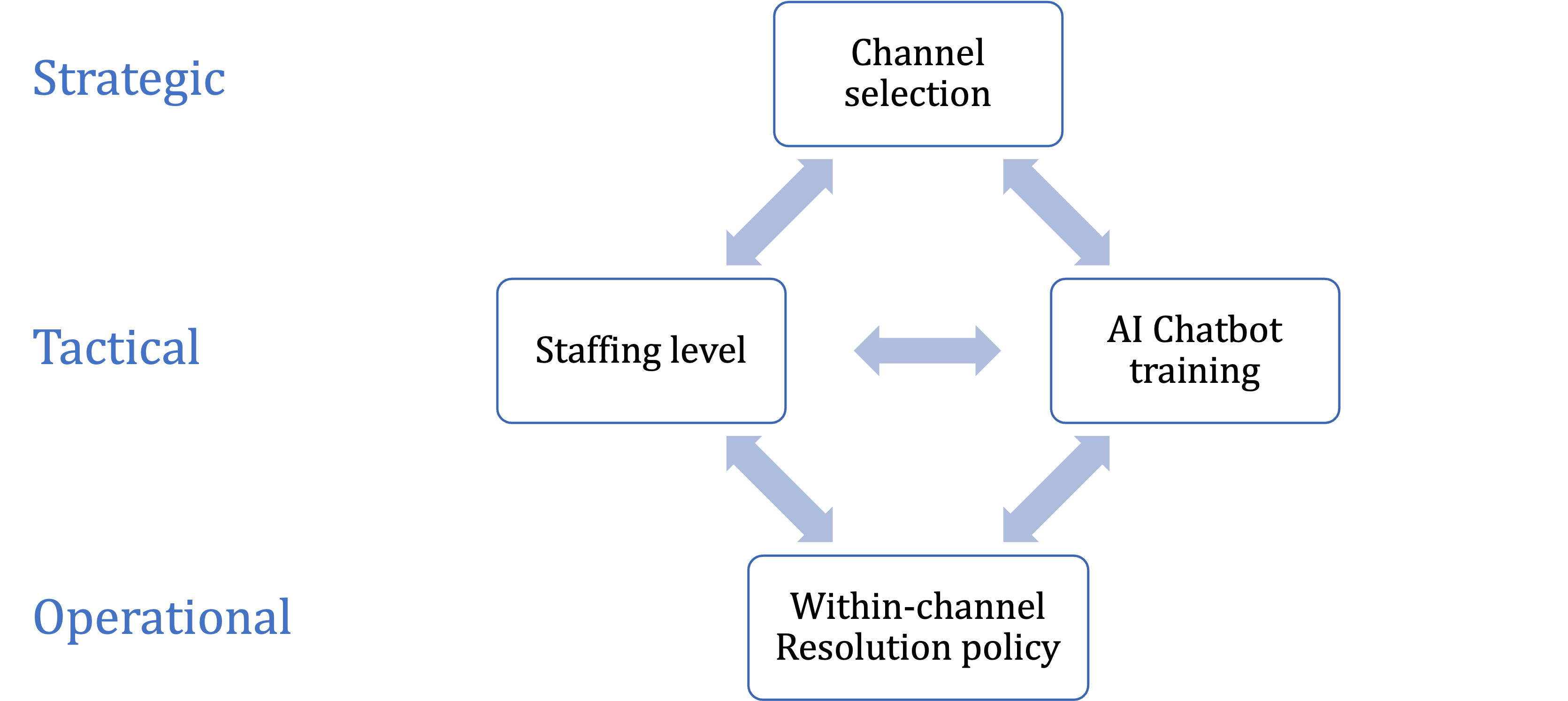}
\end{center}
\end{figure}

This paper develops analytical tools for designing and managing customer service operations. While this area is potentially vast, the relevant questions can be broadly divided into strategic (which customer service channels should be offered?), tactical (what is the right configuration of the offered channels?), and operational (what is the protocol for managing requests within each channel?), as shown in Figure \ref{fig:strat:tact:oper}. We next summarize how we have addressed each layer of the problem, and discuss what questions present interesting opportunities for future research.

We began our investigation by addressing a key operational question in customer service operations: what is the optimal request resolution strategy for the live-agent channel? To answer this question, we modeled each channel as a gatekeeper system. Within each channel, agents (or AI chatbots) act as gatekeepers receiving a stream of incoming customer requests. Requests are characterized by an ordered list of potential solutions, where each solution has a known probability of success and an associated service time. By imposing appropriate terminal conditions to induce stationarity, our formulation accommodates both limited operating hours (e.g., urgent care clinics) and 24/7 operations (e.g., airline reservation desks). We showed that although the number of admissible policies can grow exponentially, a simple structure emerges such that there are at most four optimal decision rules across the resolution steps.

To develop practical insights, we extended our analysis in two directions. We showed numerically that a heuristic threshold policy approach performs well and found a sufficiency condition that ensures its optimality. Threshold policies are computationally efficient with polynomial run time and are intuitive enough that practitioners can readily implement them. Furthermore, to capture queueing dynamics, we extended our model to include a finite waiting room. Despite the expanded policy space, we showed numerically that a simple heuristic of transferring customers when the waiting room is half full performs near-optimally.

To address the tactical challenges of determining the extent to which each channel is staffed and/or trained, and how much to invest into an AI Chatbot, we extended our operational model to incorporate the specifics of AI chatbots (near-infinite scalability and inability to warm transfer) and multiple live agents. Answering these tactical questions requires considering not only the direct costs of administering each channel, but also understanding how these choices affect customer demand. Further, these tactical questions interplay with the strategic service channel selection decision, such that the firm may need to alter its operational protocols once a new channel is introduced. We explored numerically a variety of optimal designs, which vary in channel architecture (live-agent-only, chatbot-only, both), live-agent staffing, chatbot capability, and resolution policies. Among other insights, this numerical exercise showed that, contrary to conventional wisdom, introducing an AI chatbot into a service design may increase (rather than diminish) service quality.

Our work opens several promising avenues for future research. At the operational level, our model could be extended to incorporate sophisticated routing protocols where the firm actively directs customers to specific agents (based on skill level or availability) or to specific channels based on request characteristics. This would transform our current framework, which relies on round-robin assignment within the live channel and customer self-selection across channels, into a more general matching problem that could potentially improve system performance. At the tactical level, the model could be enriched to consider how AI-assist technologies (beyond standalone chatbots) affect channel staffing and training. Such an extension would help operations managers understand the complementarities between human and AI capabilities, particularly as firms deploy technologies like real-time suggestion systems and automated documentation tools.

\bibliographystyle{pomsref}
\bibliography{mybib}

\end{document}